# SeisDiff-intp: a unified prompt-guided flow matching framework for multi-tasks seismic interpretation

Donglin Zhu, Peiayo Li, and Ge Jin

## ABSTRACT

The increasing demand for deep learning in seismic interpretation has highlighted significant challenges, particularly the reliance on massive labeled datasets and the inefficiency of training isolated models for individual tasks. To address these limitations, we introduce a unified, prompt-guided flow-matching framework capable of executing multiple seismic interpretation tasks within a single model. By conditioning on varying prompts, the model dynamically switches between interpretation objectives without requiring structural modifications. Furthermore, to overcome the scarcity of labeled data for complex subsurface features, we propose an integrated generative augmentation strategy. By employing the flow matching setting, the framework can synthesize diverse and geologically realistic training pairs, specifically targeting structurally complex. Experimental results demonstrate that the proposed approach, coupled with generative augmentation, delivers high-quality, task-specific interpretations with stable and reproducible inference behavior. Ultimately, this approach provides a scalable, flexible, and robust alternative to single-task deep learning based seismic interpretation models.

## INTRODUCTION

Seismic interpretation has long relied on extensive human expertise to identify subsurface geological structures from vast volumes of reflection data, a process that is both time-consuming and prone to interpreter inconsistency (Araya-Polo et al., 2018). Over the past decade, convolutional neural networks (CNNs) have transformed this workflow by automating key tasks including fault detection (e.g. Wu and Hale, 2016), seismic facies classification (e.g. Alaudah et al., 2019), and geobody identification (e.g. Ovcharenko et al., 2019), substantially reducing manual effort and improving repeatability across large datasets. Despite these successes, CNN-based approaches carry well-documented limitations: they are task-specific by design, require large labeled training sets, are prone to overfitting, and their inductive biases cause significant generalization degradation when applied to seismic data from surveys other than those used for training (Aminzadeh et al., 2018; Samek et al., 2017). Addressing label scarcity has motivated data augmentation strategies, semi-supervised learning, and physics-informed models (Peters et al., 2019), yet the fundamental constraint that a separate model must be trained for each interpretation objective remains unresolved.

Generative models offer a complementary pathway by learning the underlying data distribution rather than a fixed discriminative boundary. Generative adversarial networks (GANs) have been applied to seismic data augmentation and noise suppression (Laloy et al., 2018; Goodfellow et al., 2014), but their adversarial training is susceptible to instability and mode collapse, producing only a partial coverage of the target distribution (Sohl-Dickstein et al., 2015). Diffusion probabilistic models (DDPMs; Ho et al., 2020) avoid these pathologies by optimizing a likelihood-based denoising objective, which yields more stable training and broader distributional coverage. Their ability to recover clean data from noise-corrupted inputs is particularly well matched to seismic data, which are inherently noisy and often affected by acquisition footprints and incomplete illumination (Dong et al., 2021), and has been exploited for seismic denoising and reconstruction (Durall et al., 2022; Zhu et al., 2023). However, standard DDPMs require a large number of stochastic inference steps and offer no explicit mechanism for semantic task conditioning, limiting their direct applicability to multi-task seismic interpretation. Flow matching (Lipman et al., 2023; Albergo and Vanden-Eijnden, 2023) addresses the efficiency limitation by learning a deterministic continuous transport map that connects a Gaussian prior to the data distribution, eliminating the need to simulate a noisy forward diffusion process and enabling high-quality generation in far fewer steps. The rectified flow variant (Liu et al., 2022) further enforces straighter transport trajectories, producing deterministic and reproducible outputs that are especially desirable for operational seismic workflows where geological continuity must be preserved.

The emergence of geophysical foundation models represents a further step toward general-purpose seismic interpretation. Sheng et al. (2025) pre-train a Transformer-based Seismic Foundation Model (SFM) on over 2.28 million 2D seismic images from 192 global surveys, demonstrating broad generalization across classification, segmentation, inversion, denoising, and interpolation tasks. However, SFM still requires assembling large task-specific labeled datasets and training a dedicated decoder for each downstream objective. Gao et al. (2026) propose SAG, a promptable foundation model for universal geobody interpretation that uses a single prompt-conditioned decoder to segment multiple geobody types across surveys without per-task retraining. Despite these advances, SAG still depends on large annotated training datasets and accepts only geometric prompts (points, bounding boxes, and well logs), with no support for free-form natural-language task descriptions.

This work introduces SeisDiff-intp, a prompt-guided flow-matching algorithm for unified multi-task seismic interpretation (Figure 1). SeisDiff-intp addresses the above limitations by combining two complementary capabilities within a single generative framework: a rectified flow model conditioned on natural-language prompts that switches between geological interpretation objectives at inference time without any retraining, and a generative augmentation strategy that repurposes the same flow matching framework to synthesize diverse, geologically realistic labeled training pairs for data-scarce targets such as mass-transport deposits. Together, these capabilities provide a scalable and flexible model for multi-tasks seismic interpretation.

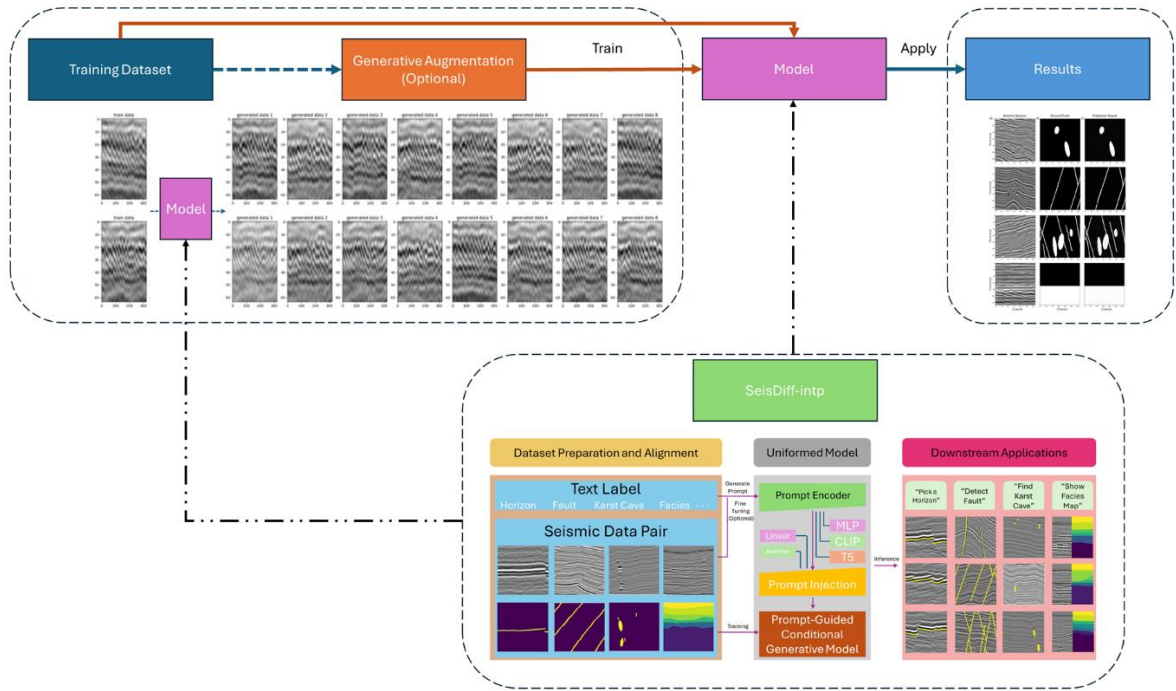

Figure 1. The proposed workflow.

## METHOD

We propose SeisDiff-intp (Figure 2), a prompt-guided generative framework for seismic interpretation that supports multiple tasks such as fault detection, geobody delineation, and seismic facies classification within a single unified model. Seismic interpretation is formulated as a segmentation problem in which the seismic image serves as a fixed conditioning input and the segmentation mask is treated as the generative variable.

Semantic task control is achieved by encoding prompts using a dedicated prompt encoder by employing classifier-free guidance to incorporate guidance information during the generative process. This design allows the model to dynamically adjust interpretation objectives at inference time without retraining or changing network structure, improving both scalability and interpretability.

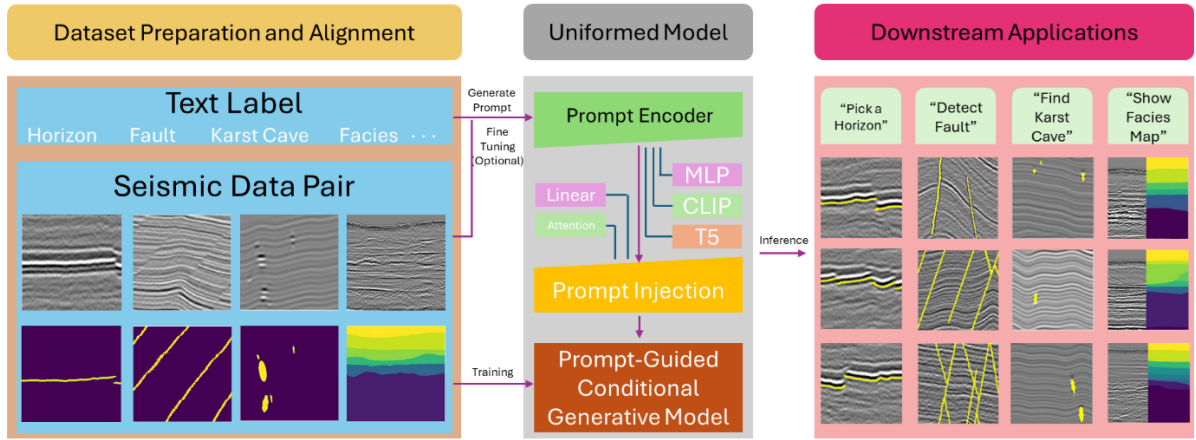

Figure 2. The SeisDiff-intp model

**Background**

*Denoising Diffusion Probabilistic Models*

Denoising diffusion probabilistic models (DDPMs; Ho et al., 2020) are generative models that learn a data distribution by progressively adding Gaussian noise to clean samples and then learning to reverse this process. In the context of seismic interpretation, DDPMs can model complex subsurface structures by learning the statistical distribution of seismic images or interpretation masks.

Let $x_0$ denote a clean data sample, such as a seismic image or segmentation mask. The forward diffusion process generates a noisy sample $x_t$ at time step $t$ according to

$$q(x_t \mid x_{t-1}) = \mathcal{N}\left(x_t; \sqrt{1-\beta_t}\, x_{t-1}, \beta_t I\right), \tag{1}$$

where $\mathcal{N}(\cdot)$ denotes a Gaussian distribution, $I$ is the identity matrix, and $\beta_t \in (0,1)$ controls the variance of the injected noise at time step $t$.

The reverse diffusion process is parameterized as

$$p_\theta(x_{t-1} \mid x_t) = \mathcal{N}(x_{t-1}; \mu_\theta(x_t, t), \Sigma_\theta(x_t, t)), \tag{2}$$

where $\mu_\theta$ and $\Sigma_\theta$ denote the mean and covariance predicted by a neural network with parameters $\theta$.

Training minimizes a variational lower bound, but in practice a simplified noise-prediction loss is commonly used,

$$\mathcal{L}_{\text{simple}} = \mathbb{E}_{x_0, \epsilon, t}\left[\|\epsilon - \epsilon_\theta(\sqrt{\bar{\alpha}_t}\, x_0 + \sqrt{1-\bar{\alpha}_t}\, \epsilon, t)\|_2^2\right], \tag{3}$$

where $\epsilon \sim \mathcal{N}(0, I)$ is Gaussian noise, $\epsilon_\theta$ is the network's noise prediction, $\alpha_t = 1 - \beta_t$, and $\bar{\alpha}_t = \prod_{i=1}^{t} \alpha_i$.

*Improved DDPMs and DDIMs*

Improved DDPMs (Nichol and Dhariwal, 2021) introduce learned variance and a cosine noise schedule to improve sample quality and convergence. The hybrid objective combines the simplified loss with a variational term,

$$\mathcal{L}_{\text{hybrid}} = \lambda \mathcal{L}_{\text{vlb}} + \mathcal{L}_{\text{simple}}, \tag{4}$$

where $\lambda$ controls the contribution of the variational loss.

The cosine noise schedule is defined as

$$\bar{\alpha}_t = \frac{f(t)}{f(0)}, \tag{5}$$

$$f(t) = \cos^2\left(\frac{t/T+s}{1+s} \cdot \frac{\pi}{2}\right), \tag{6}$$

where $T$ is the total number of diffusion steps and $s$ prevents excessive information loss at early timesteps.

Denoising diffusion implicit models (DDIMs; Song et al., 2021) accelerate inference by defining a non-Markovian reverse process,

$$x_{t-1} = \sqrt{\alpha_{t-1}} \frac{x_t - \sqrt{1-\alpha_t}\,\epsilon_\theta(x_t,t)}{\sqrt{\alpha_t}} + \sqrt{1-\alpha_{t-1}-\sigma_t^2}\,\epsilon_\theta(x_t,t) + \sigma_t \epsilon, \tag{7}$$

where $\sigma_t$ controls the level of stochasticity and $\epsilon \sim \mathcal{N}(0, I)$.

*Flow matching and rectified flow*

Flow matching reformulates generative modeling as learning a deterministic transport between distributions (Lipman et al., 2023). Given a clean sample $x_0$ and a noise sample $x_1 \sim \mathcal{N}(0, I)$, an interpolated state is defined as

$$x_t = (1-t)x_0 + tx_1, t \in [0,1]. \tag{8}$$

The corresponding target velocity field is

$$v(x_t, t) = \frac{dx_t}{dt} = x_1 - x_0. \tag{9}$$

A neural network $v_\theta$ is trained by minimizing

$$\mathcal{L}_{\text{FM}} = \mathbb{E}_{x_0, x_1, t}[\| v_\theta(x_t, t) - (x_1 - x_0) \|_2^2]. \tag{10}$$

Rectified flow (Liu et al., 2023) further improves training stability and sample quality by enforcing straighter transport paths. Generation is performed by integrating the ordinary differential equation

$$\frac{dx_t}{dt} = v_\theta(x_t, t), \tag{11}$$

which allows high-quality outputs to be generated in only a small number of steps. This deterministic property is particularly beneficial for seismic interpretation, where preserving geological continuity is critical.

**SeisDiff-intp methodology**

*Prompt-guided generative modeling with classifier-free guidance*

Standard diffusion and flow-matching models lack explicit semantic control, making them unsuitable for multi-task seismic interpretation. Prompt-guided generative modeling addresses this limitation by conditioning the generative process on textual prompts.

In SeisDiff-intp, prompts are encoded into fixed-length embeddings using a prompt encoder trained jointly with the generative model. Rather than injecting prompts fully through cross-attention or direct feature concatenation, prompt conditioning is applied exclusively through classifier-free guidance.

During training, the model is exposed to both conditional samples, where the text embedding is provided, and unconditional samples, where the text embedding is replaced with a null token. During inference, conditional and unconditional predictions are combined as

$$x_t = f(x_t \mid \text{prompts}) + (1 - \omega)f(x_t \mid \emptyset), \tag{12}$$

where $f(\cdot)$ denotes the predicted results, and $\omega$ controls the strength of prompts guidance. This formulation allows continuous adjustment of prompt influence at inference time without retraining the model.

*Network architecture*

**Multi-Scale Diffusion Transformer (DiT) Architecture.** In this study, we adopt and modified a multi-scale DiT (Peebles and Xie, 2022) that operates in token space while preserving the hierarchical feature learning paradigm of encoder–decoder networks (Figure 3). Given an input seismic image of size 128×128, the model first applies a patch embedding layer with a patch size of 16, resulting in an 8×8 grid of tokens. Each token is projected into a latent embedding space, forming the input sequence to the transformer backbone.

The DiT backbone is organized into multiple stages following an encoder–bottleneck–decoder structure. In the configuration used in this work, three stages are employed. The encoder stage operates on the fine token grid with an embedding dimension of 256 and a depth of eight transformer blocks. Tokens are then merged via a token-space downsampling operation to form a coarser 4×4 representation, which is processed by a bottleneck stage with an embedding dimension of 512 and eight transformer blocks. Subsequently, tokens are expanded back to the original resolution and fused with encoder features through skip connections, followed by a decoder stage that mirrors the encoder in both embedding dimension and depth.

Multi-scale feature learning is realized by explicit token merging and expansion operations between stages, analogous to spatial downsampling and upsampling in general end-to-end architectures. Skip connections between corresponding encoder and decoder stages are used to preserve high-resolution information, enabling the model to jointly capture fine-scale details and large-scale structural context. Unlike pixel-space in CNNs like UNets, this hierarchy is constructed entirely in token space, allowing global self-attention to be applied at each stage.

Each transformer block employs adaptive layer normalization (AdaLN), in which the normalization parameters are modulated by a conditioning vector derived from the diffusion time step and optional class labels. This design enables classifier-free guidance by randomly dropping class conditioning during training and setting the label to a null token during unconditional inference. To encode spatial relationships among tokens, two-dimensional axial rotary positional embeddings (RoPE) are applied to the query and key vectors within the self-attention mechanism. RoPE provides relative positional encoding and translation equivariance, which is particularly advantageous for tiled inference and seismic data characterized by spatial continuity.

All architectural hyperparameters, including the number of transformer stages, embedding dimensions, number of attention heads, and transformer depth at each stage, are explicitly specified through an external configuration and summarized in Table 1.

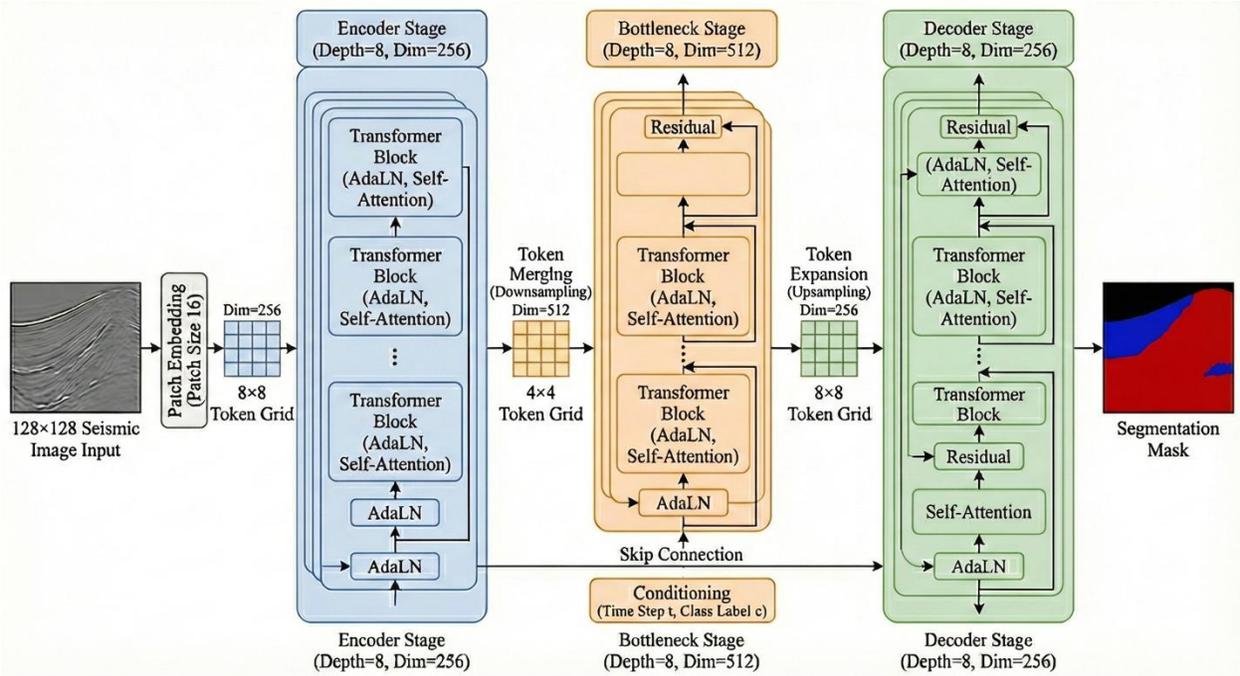

Figure 3. DiT architecture.

**Conditioning and task control.** Temporal conditioning is introduced through diffusion or flow time embeddings that encode the current generative timestep $t$. These embeddings are injected into every transformer block using adaptive layer normalization (AdaLN), allowing the network to adjust its behavior dynamically along the generative trajectory.

The task conditioning is achieved through textual prompts that describe the interpretation objective, such as faults, karst collapse, or mass-transport deposits. The prompts are encoded into fixed-length embeddings using a text encoder, and the resulting embeddings are combined with the time embedding to form the conditioning vector for the transformer backbone. By conditioning the model on prompts rather than task-specific output heads, the proposed framework supports multiple interpretation tasks within a single unified model, allowing task switching at inference time without modifying network weights. As a result, SeisDiff-intp supports multiple interpretation tasks within a single unified model.

*Training with rectified flow and high-resolution adaption*

SeisDiff-intp is trained using a rectified flow (RF) formulation rather than DDPM or DDIM objectives. In this framework, the generative task is formulated as learning a deterministic velocity field that transports a noisy segmentation mask toward its clean ground-truth counterpart under seismic conditioning.

Let $x_0$ denote the ground-truth segmentation mask and $x_1 \sim \mathcal{N}(0, I)$ denote a Gaussian noise sample. An intermediate state $x_t$ is constructed along a linear interpolation path,

$$x_t = (1 - t)x_0 + tx_1, \tag{13}$$

where $t \sim \mathcal{U}(0,1)$ is a continuous time variable. The target velocity field is defined as $x_1 - x_0$, and the network $v_\theta$ is trained to predict this velocity conditioned on the seismic image $I$ and the text prompt $D$,

$$\mathcal{L}_{\text{RF}} = \mathbb{E}_{x_0, x_1, t}[\| v_\theta(x_t, I, D, t) - (x_1 - x_0) \|_2^2]. \tag{14}$$

This formulation avoids variance scheduling and stochastic noise sampling during inference, resulting in a simpler and more stable training objective. Because the learned flow is deterministic, inference is repeatable and produces consistent interpretation results, which is desirable for operational seismic workflows.

Although the model is trained primarily on fixed-size seismic patches, seismic interpretation is typically performed on large-scale sections that contain long-range structural information. To reduce the resolution gap between training and deployment, a high-resolution adaptation (HRA) phase is introduced at the final stage of training (Touvron et al., 2019; Oquab et al., 2023).

After the model converges on 128×128 patches, training is continued for a limited number of epochs using larger spatial crops while reducing the learning rate. This adaptation allows the transformer backbone to adjust its attention statistics and multi-scale interactions to broader spatial contexts without modifying the network architecture or retraining from scratch. Importantly, the HRA phase preserves local structural priors learned during earlier training stages while improving global consistency during large-scale inference.

*Inference and fast sampling*

During inference, segmentation masks are generated by integrating the learned velocity field from $t = 1$ to $t = 0$ using a small number of deterministic ODE solver steps (e.g., Euler):

$$x_{t+\Delta t} = x_t + \Delta t v_\theta(x_t, I, D, t). \tag{15}$$

Because the rectified flow formulation is deterministic, no stochastic noise injection or multiple realizations are required. This results in fast sampling suitable for large seismic volumes, stable and repeatable outputs, and clean, geologically consistent segmentation masks.

Classifier-free guidance is applied during inference by combining conditional and unconditional velocity predictions. The guidance scale controls the strength of semantic conditioning, allowing interpreters to adjust task specificity without retraining.

Table 1 Hyper parameters

| Hyper Parameters | |
|---|---|
| Epoch | 5000 |
| Batch Size | 12 |
| Learning Rate | 0.0001 |
| Patch Size | 16 |
| Block Number | 8 |
| Head Number | 8 |
| Embedding Dimension | [256, 512] |
| Training Sample Number | 6360 |
| Training Sample Input Size | 128 x 128 |
| Sampling Method | ODE |
| ODE Solver | Euler |
| Sampling Steps | 20 |

**Multi-tasks training dataset construction**

To train SeisDiff-Intp for unified seismic interpretation, a multi-task dataset is constructed consisting of seismic images, segmentation masks, and corresponding text prompts. The dataset includes three primary interpretation targets: faults, karst collapse features, and mass-transport deposits. Each target type is paired with a descriptive text prompt, enabling the model to learn the association between semantic conditions and geological expressions. The samples with mixed features (both fault and karst collapse) are also included in the training and testing dataset.

*Fault and karst Collapse synthetic dataset*

The karst collapse dataset contains seismic images with void-like or disrupted reflectors, representing karst collapse features. Variations in size, depth, and reflectivity contrast are included to challenge the model's ability to distinguish subtle geobodies from layered backgrounds. These examples emphasize geobody segmentation under weak amplitude contrast.

The fault dataset consists of synthetic seismic sections with faulted structures following the methodology of Wu et al. (2018). These datasets include variations in fault orientation, displacement, and spatial density, while providing pixel-accurate segmentation masks. The diversity of fault geometries helps the model learn robust fault representations under different structural scenarios.

*Mass-transport deposit dataset with generative augmentation*

In seismic reflection data, mass-transport deposits (MTDs) typically appear as internally chaotic to semi-transparent packages bounded by basal shear surfaces and erosional tops, and they commonly display structural partitioning into updip extensional domains and downdip contractional toes (Posamentier and Martinsen, 2011; Moscardelli and Wood, 2016). In the study area, the target mass-failure bodies exhibit a well-developed downdip contractional domain characterized by imbricate thrust faults and associated folding, a deformation style widely documented in submarine landslides and MTD toe regions (Alsop and Holdsworth, 2017). Although carbonate MTDs are often regarded as mechanically heterogeneous and unfavorable reservoir intervals, subsurface and outcrop studies from the Delaware Basin demonstrate that mixed carbonate–siliciclastic mass-transport complexes, such as those within the Delaware Mountain Group, can exhibit strong structural compartmentalization and localized enhancement of fracture-controlled permeability, particularly within contractional toe domains (Smye et al.,

2021; Simabrata et al., 2025). Accurate seismic delineation of these structurally complex features is therefore critical for reducing interpretation uncertainty and for assessing their potential impact on reservoir performance and development decisions.

The availability of accurately labeled MTDs examples in field seismic data is typically limited, which poses a significant challenge for data-driven seismic interpretation. To address this issue, we develop a compositional dataset construction workflow that generates realistic seismic–MTDs training pairs by reusing the same rectified flow formulation adopted in SeisDiff-Intp, but in an unconditional generative setting. This approach enables the synthesis of diverse, geologically faithful MTD realizations without relying on manual drawing or simplified synthetic templates.

We first extract representative MTDs patches from field seismic surveys. These samples preserve authentic morphological and textural characteristics of MTDs, including chaotic internal reflections, disrupted bedding, basal erosion surfaces, and headwall scarps. Each extracted MTDs patch is treated as a target realization, denoted by $\mathbf{x}_0$, which represents a clean MTDs example in data space. In contrast to the interpretation task, no seismic conditioning is applied during this stage. Instead, the generative model is trained to transform random noise into MTDs structures, allowing the learned flow to capture the intrinsic geometry of MTDs independent of background stratigraphy.

When using the deterministic rectified flow formulation alone, the learned transport map produces a unique trajectory for each noise realization. In practice, this results in generated MTDs samples (Figure 4) that concentrate around representative training exemplars, yielding limited morphological diversity. While such determinism is desirable for seismic interpretation, where stable and reproducible outputs are critical, it is insufficient for dataset construction, where broad structural variability is required to improve model generalization.

To enrich diversity during data synthesis, additional stochastic perturbations are injected at intermediate time steps along the rectified trajectory according to

$$\mathbf{x}_t \leftarrow \mathbf{x}_t + \sigma(t)\epsilon, \tag{16}$$

where $\epsilon \sim \mathcal{N}(0, I)$ is Gaussian noise and $\sigma(t)$ is a time-dependent noise amplitude. In this work, $\sigma(t)$ is defined as

$$\sigma(t) = \sigma_{\max} t(1-t), \tag{17}$$

where $\sigma_{\max}$ (set to 0.15 in this study) is a predefined scalar that controls the maximum noise strength. This formulation ensures that the injected noise peaks at intermediate times ($t \approx 0.5$) and vanishes at the endpoints ($t = 0$ and $t = 1$), thereby preserving the boundary conditions imposed by the rectified flow. This operation encourages exploration of broader MTDs manifold while maintaining geological plausibility, enabling the generation of MTDs realizations with substantial variability in geometry, scale, orientation, and internal texture.

In parallel, clean background seismic sections are extracted from field data to provide realistic stratigraphic contexts characterized by continuous reflectors and natural noise patterns. Final training pairs are constructed by compositing a flow-generated MTDs realization into a background seismic section. Boundary smoothing is applied to ensure physically plausible transitions between the MTDs and surrounding strata. The spatial footprint of the inserted MTDs is recorded as a binary segmentation mask, yielding paired seismic images and label maps suitable for supervised training.

This compositional workflow produces a large collection of high-fidelity, label-accurate seismic–MTDs pairs that combine realistic seismic backgrounds with diverse, flow-generated

MTD expressions. By employing RF for interpretation and dataset synthesis, the proposed approach exploits the complementary strengths of both formulations. The resulting dataset exhibits substantial geological variability and structural realism, significantly strengthening the robustness and generalization capability of SeisDiff-intp.

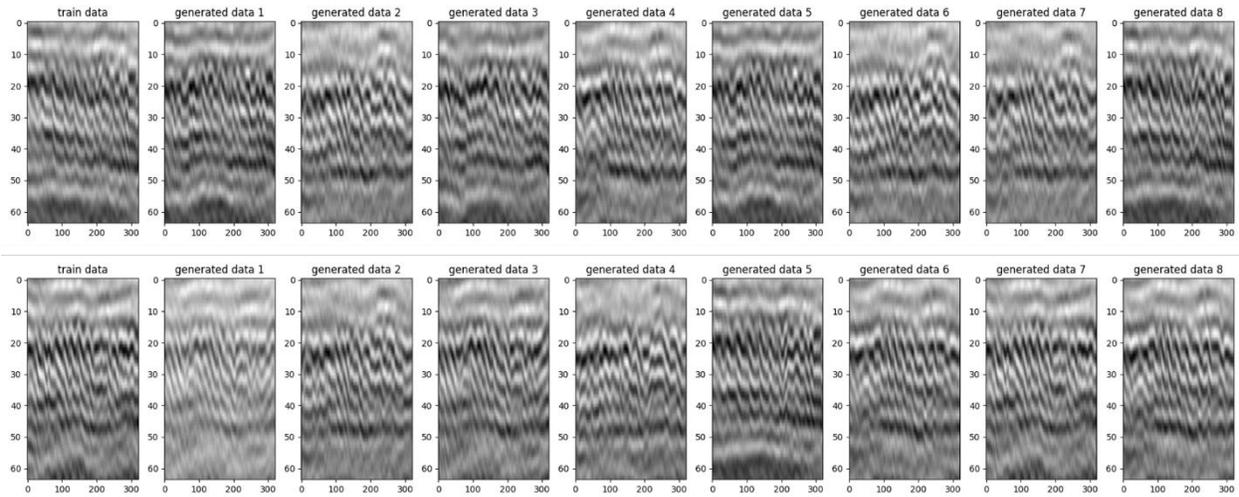

Figure 4. Examples of generated MTDs. The left column is the training data, the rest are generated samples.

## APPLICATION

To demonstrate the versatility of SeisDiff-intp, we apply the model to multiple seismic interpretation tasks, including fault detection, karst collapse identification, and MTDs delineation. These tasks span a range of structurally complex and geologically diverse interpretation objectives. The model is trained using a dataset that includes simulated faults, synthetic karst features, and generated MTDs, enabling the network to learn generalized structural and stratigraphic representations rather than task-specific patterns.

We investigate the model's ability to distinguish mixed geological features, which are far more commonly observed in field seismic data than isolated structural elements. A representative example is a paleo-karst system, which typically comprises a combination of faults, karst collapses, and karst caves. To evaluate this scenario, we use a synthetic testing dataset containing coexisting fault and karst-collapse features within a seismic vertical section (Figure 5). By applying different prompts, "Find both faults and karst collapse", "Detect faults", "Pick karst collapse", and "Detect MTDs", the guided model produces distinct and target-specific responses for each interpretation objective. In contrast, single-task models exhibit misclassification, blurred boundaries, and increased false positives when applied to mixed-feature scenarios (Figure 6). These results demonstrate that a unified, prompt-guided multi-task model can provide more accurate and efficient seismic interpretation than multiple independently trained single-task models.

We apply the trained model to post-stack time-migrated seismic data from the Fasken Ranch area in the Permian Basin, West Texas, where carbonate slope-to-basin systems commonly contain carbonate-rich MTDs. Figures 7(a) and Figure 8(a) show a representative seismic time slice and vertical section containing MTDs, with particular emphasis on the contractional toe domain. Using the prompt "Find MTDs", the detection results from proposed method are overlaid on the seismic data in Figure 7(b) and Figure 8(b). In the seismic section, the predicted regions successfully delineate MTD characteristics, including chaotic reflections and imbricate subtle thrust-fault seismic response. The time-slice results ($t = 1.35s$) show good spatial agreement with manually interpreted boundaries of the MTD contractional toe zone.

Moreover, we conduct an additional experiment in which generated MTDs are excluded and the model is trained directly using a limited set of manually labeled MTD examples from field data shown in Figure 7(c) and Figure 8(c). Although the model is still capable of identifying

MTDs-related features under this setting, the results contain a significantly higher number of false positives, highlighting the importance of generative augmentation for robust and stable MTDs delineation.

In addition, the trained model is applied to a fault detection task to evaluate its performance under different prompt guidance. For fault interpretation, a simple semantic prompt such as "Detect faults" is sufficient to steer the model toward generating coherent fault segmentation masks. Figure 9(a) shows a representative seismic vertical section from the F3 dataset. The corresponding prediction from the proposed model (Figure 9(b)) demonstrates strong fault continuity. For comparison, Figure 9(c) presents result from a non-generative baseline model (UNet) which is trained solely on fault examples from the same training dataset of proposed method. While both approaches yield comparable overall fault detection performance on major faults, the proposed generative model shows improved sensitivity to potential existing secondary and subtle fault structures, indicating enhanced generalization and structural awareness. When an unrelated prompt such as "Segment karst collapse" is applied, the model produces only sparse false-positive responses (Figure 9(d)), further confirming the effectiveness of semantic prompt guidance in controlling the model's interpretational focus.

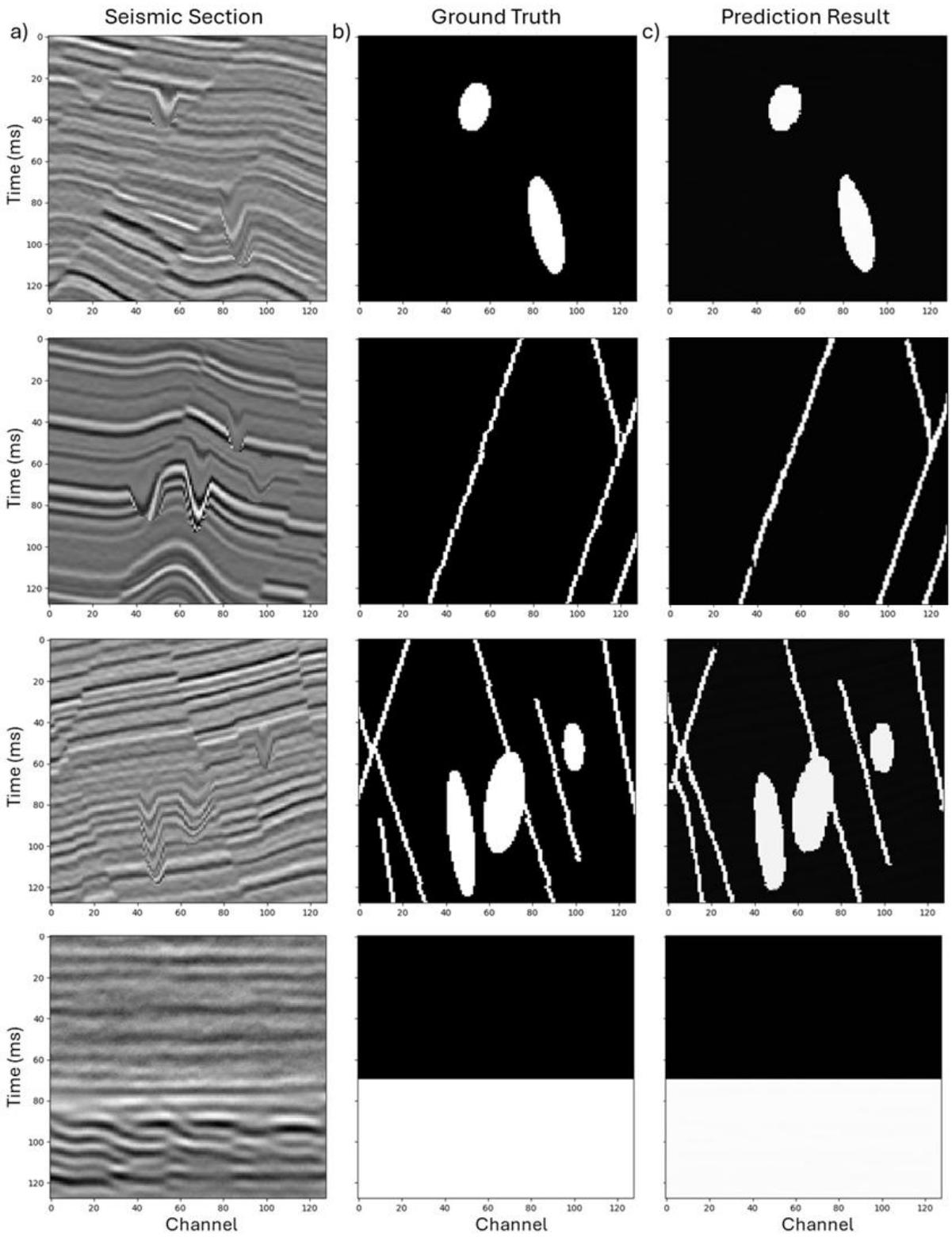

Figure 5. Synthetic detection results using various prompts. The columns display: (a) the input seismic section, (b) the ground truth, and (c) prediction results from the proposed method. The rows correspond to the following prompts (from top to bottom): 'karst', fault', 'karst and fault', and 'MTDs'.

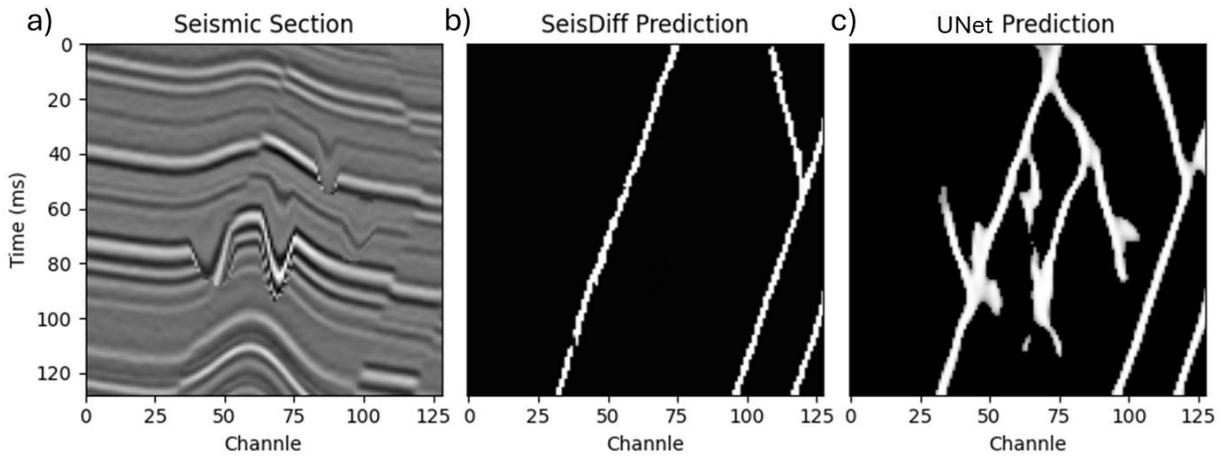

Figure 6. Comparison of synthetic detection results, (a) seismic section containing fault and karst features, (b) fault detection result using the proposed method, (c) fault detection result using UNet.

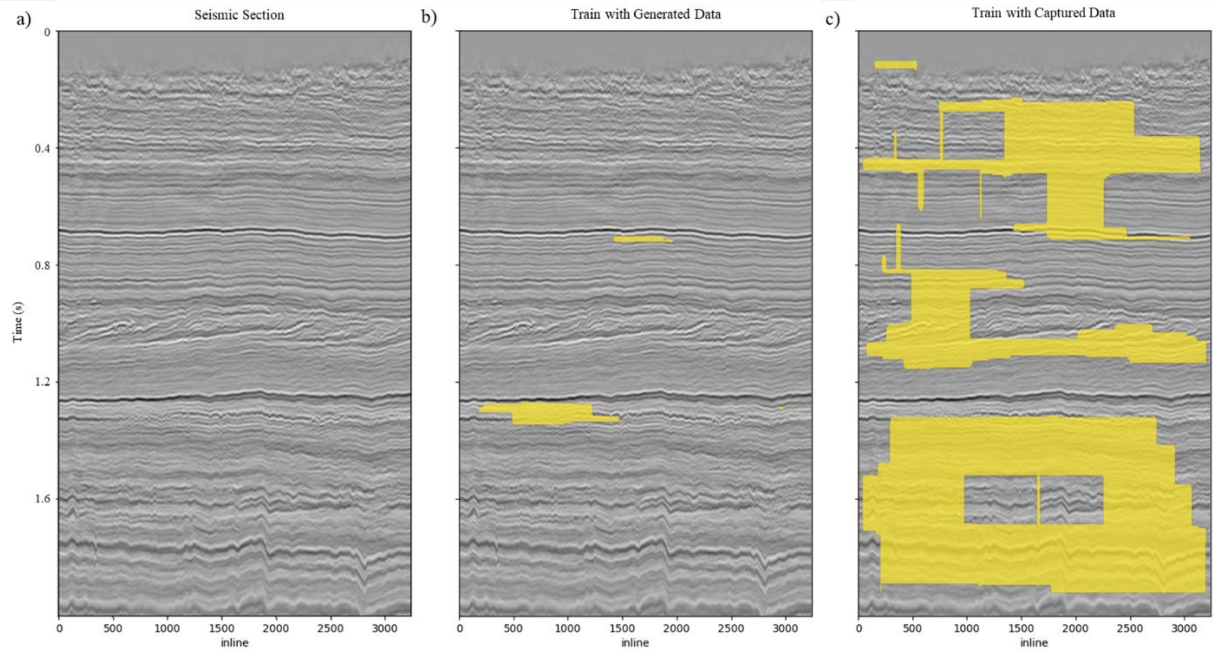

Figure 7. Seismic section with MTDs detection results, a) input seismic section, b) MTDs detection mask trained by generated data overlaid on seismic section, and c) MTDs detection mask trained by manually labeled field data overlaid on seismic section

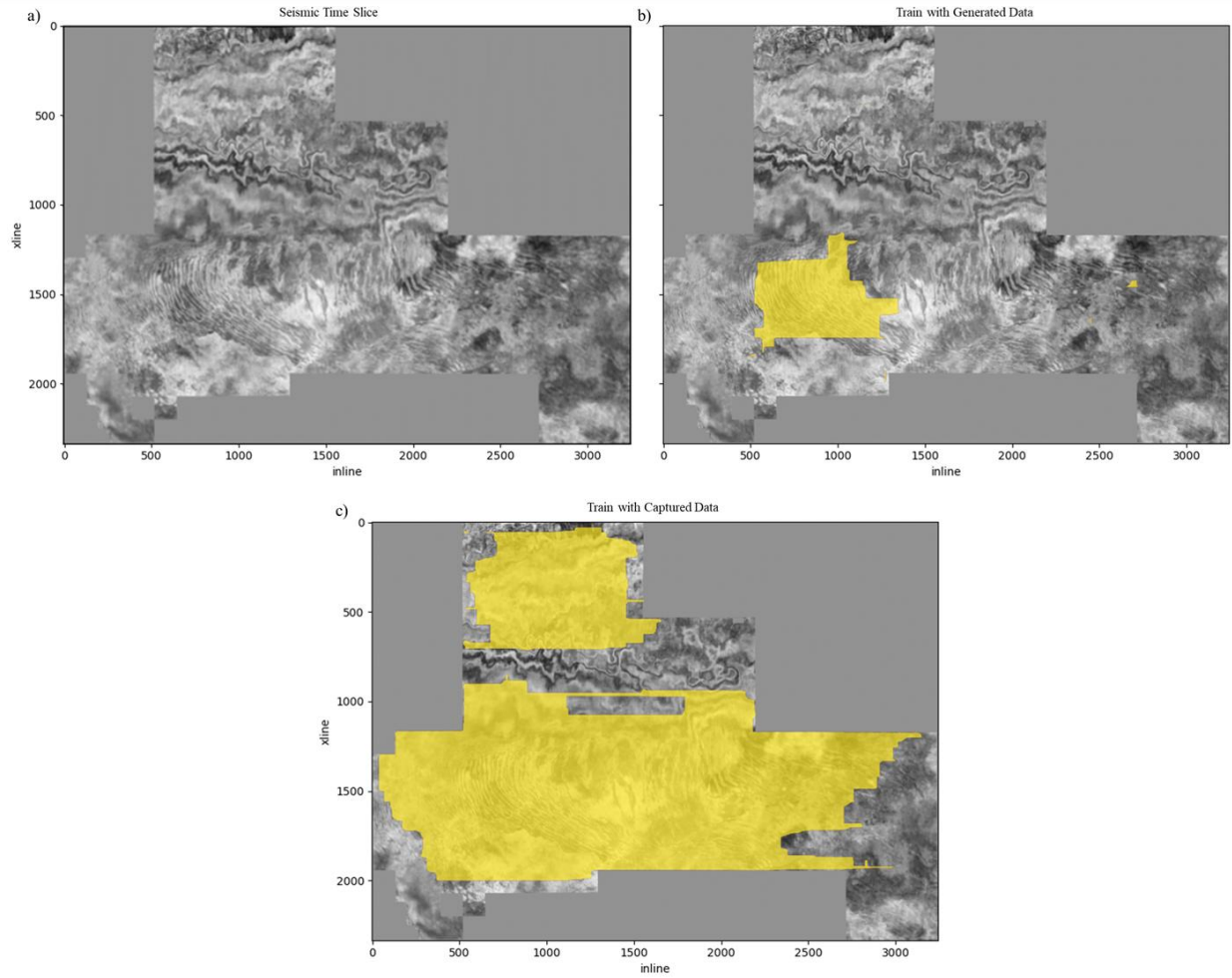

Figure 8. Seismic time slice with MTDs detection results, a) seismic time slice, b) MTDs detection mask trained by generated data overlaid on seismic time slice, and c) MTDs detection mask trained by captured field data overlaid on seismic time slice

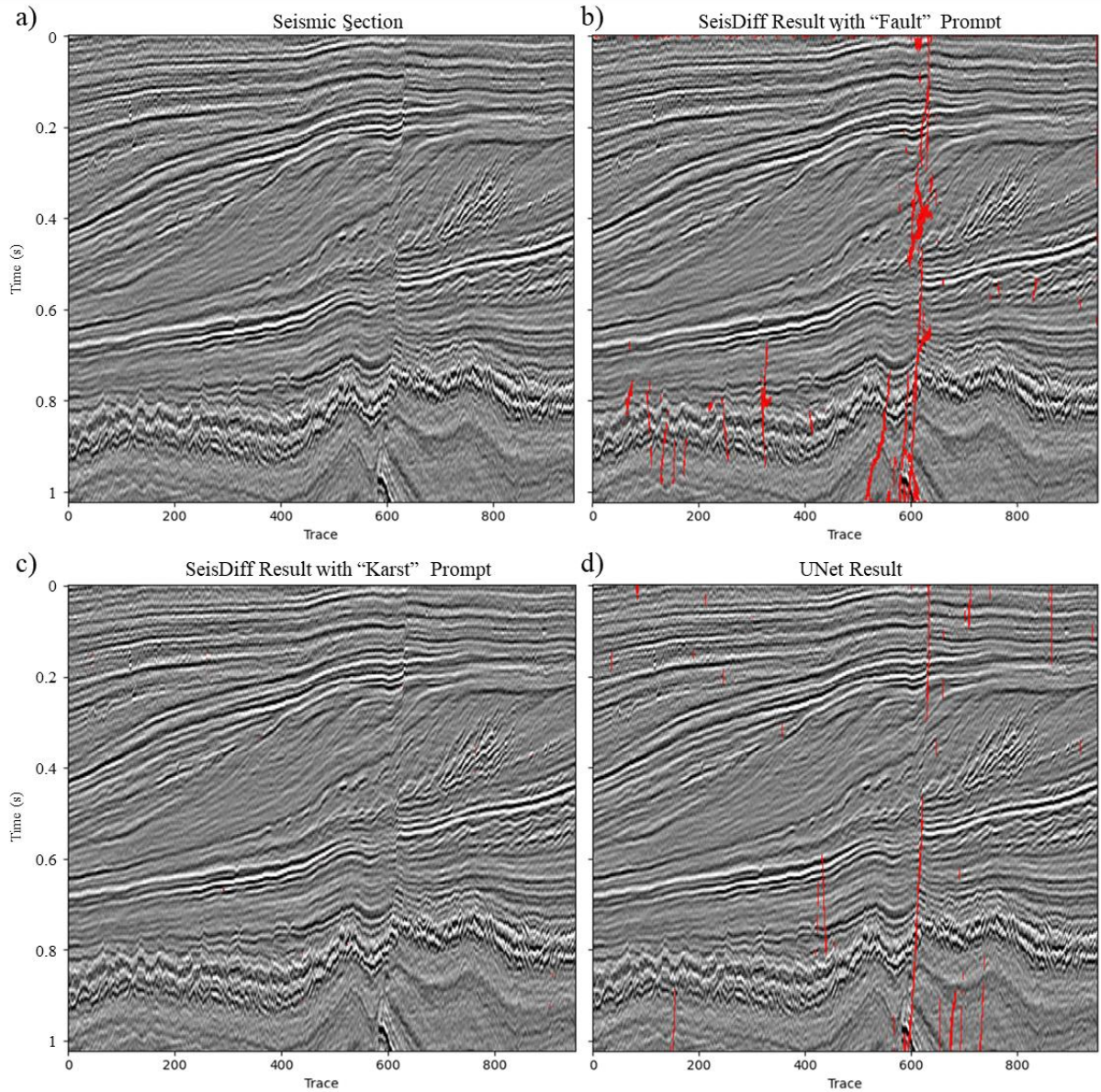

Figure 9. Seismic section with fault detection results, (a) input seismic section, (b) detection masks overlaid on the seismic section by the proposed method using the 'fault' prompt, (c) the proposed method using the 'karst' prompt, and (d) the U-Net baseline.

# DISCUSSION

*Limitations*

SeisDiff-intp has several limitations that should be acknowledged. First, the current model operates on 2D seismic sections and does not natively handle 3D seismic volumes. Extending the framework to 3D would require substantially more training data, higher GPU memory, and a redesigned patch-embedding strategy to capture along-strike structural continuity, which is not addressed in the present work. Additionally, the quality of the generative augmentation depends on the representativeness of the field seismic patches used as training samples. If the available field examples cover only a narrow range of geometries or geological settings, the diversity of the synthesized samples will be correspondingly limited, potentially reducing generalization to sample expressions not represented in the seed data. Finally, the text prompt encoder is trained jointly with the generative model on a small vocabulary of geological target descriptions. Prompts that describe geological features outside the training vocabulary may produce unreliable or poorly constrained segmentation results, and the model does not currently provide feedback to the user when a prompt is ambiguous or out of distribution.

*Future work*

While SeisDiff-intp provides a unified prompt-controlled framework for seismic interpretation, several important research directions remain open. A key area for further development lies in enhancing the depth of language understanding used for prompt conditioning. In the current implementation, the prompt encoder is a lightweight module trained jointly with the generative model on a small set of short geological target labels. As a result, the model can distinguish between broad interpretation objectives such as "detect faults" or "find MTDs", but it cannot parse or act on complex geological descriptions that require reasoning about spatial relationships, structural context, or multi-attribute constraints. A natural and promising direction is to replace the lightweight prompt encoder with a large language model (LLM) or a geoscience-domain language model, which have been pre-trained on extensive datasets and can represent nuanced semantic content. Such a language backbone would enable the model to condition on rich, multi-clause descriptions and translate that geological reasoning directly into the generative conditioning vector. This would move SeisDiff-intp from label-based task switching toward genuine language-guided geological interpretation, opening the door to interactive workflows in which interpreters communicate with the model in the same descriptive language they use with colleagues, rather than through predefined categorical prompts.

Another important direction is the integration of pre-trained foundation model encoders into the SeisDiff-intp framework. In the current implementation, the seismic image encoder within the DiT backbone is trained from scratch, which may limit the quality of seismic feature representations. A natural extension would be to replace or augment this encoder with a pre-trained seismic foundation model, whose encoder has been exposed to over two million diverse seismic images through self-supervised pre-training and has learned rich, transferable representations of subsurface structures. Incorporating such an encoder into the conditioning pathway of SeisDiff-intp would allow the generative model to benefit from large-scale seismic prior knowledge without requiring a correspondingly large labeled dataset for interpretation training. This combination would address a core tension in the current design: the flow matching framework provides powerful generative and prompt control capabilities, but its seismic feature extraction is constrained by the relatively small size of the interpretation training set. Such a hybrid architecture, a generative flow-matching backbone conditioned through a foundation model encoder, would represent a step toward a truly unified model that combines the cross-survey generalization of foundation models with the generative augmentation and language-based task control introduced in this work.

## CONCLUSION

In this study, we introduced SeisDiff-intp, a prompt-controlled flow matching framework for seismic interpretation that integrates a modified Diffusion Transformer architecture with conditional generative modeling. Unlike conventional deep learning–based segmentation approaches, which typically require separate task-specific models and retraining for different interpretation objectives, SeisDiff-intp provides a unified framework in which task adaptation is achieved through prompts conditioning rather than architectural/dataset modification.

Experimental results demonstrate that SeisDiff-intp is capable of providing high-quality seismic interpretation results across multiple tasks while maintaining stable and reproducible inference behavior. By controlling interpretation objectives through prompts instead of training multiple independent models, the proposed framework offers a flexible and scalable alternative to traditional single-task workflows. This design promotes parameter efficiency, shared representation learning, and improved consistency across interpretation tasks, representing a meaningful step toward more adaptable and interpretable seismic interpretation workflow.

In addition, this work highlights the importance of data-centric modeling in seismic interpretation. By combining generative augmentation with dataset construction strategy grounded in field seismic data, the proposed approach effectively addresses label scarcity while preserving geological realism. Together, these contributions demonstrate how generative models can be leveraged not only for interpretation, but also for scalable and robust training data construction in complex geophysical settings.